\pdfoutput=1
\documentclass{PoS}

\title{Predicting the neutralino relic density in the MSSM more precisely}
\ShortTitle{Predicting the neutralino relic density in the MSSM more precisely}

\author{Julia Harz\\
        Sorbonne Universit\'es, Institut Lagrange de Paris (ILP), 98bis Boulevard Arago, F-75014 Paris, France\\
        Sorbonne Universit\'es, UPMC Univ Paris 06, UMR 7589, LPTHE, F-75005 Paris, France\\
        CNRS, UMR 7589, LPTHE, F-75005 Paris, France}

\author{\speaker{Bj\"orn Herrmann}\\
		LAPTh, Universit\'e Savoie Mont Blanc, CNRS, 9 Chemin de Bellevue, F-74941 Annecy-le-Vieux, France}

\author{Michael Klasen, Karol Kova\v{r}\'ik, Patrick Steppeler\\
        Institut f\"ur Theoretische Physik, Westf\"alische Wilhelms-Universit\"at M\"unster, Wilhelm-Klemm-Stra{\ss}e 9, 		D-48149 M\"unster, Germany}

\abstract{The dark matter relic density being a powerful observable to constrain models of new physics, the recent experimental progress calls for more precise theoretical predictions. On the particle physics side, improvements are to be made in the calculation of the (co)annihilation cross-section of the dark matter particle. We present the project \texttt{DM@NLO} which aims at calculating the neutralino (co)annihilation cross-section in the MSSM including radiative corrections in QCD. In the present document, we briefly review selected results for different (co)annihilation processes. We then discuss the estimation of the associated theory uncertainty obtained by varying the renormalization scale. Finally, perspectives are discussed.}

\FullConference{38th International Conference on High Energy Physics\\
		3-10 August 2016\\
		Chicago, USA}

\begin{document}

\section{Introduction}

New physics models including a candidate for the cold dark matter (CDM) in our Universe can be tested by predicting the dark matter abundance $\Omega_{\chi}h^2$, 
which is the product of mass and present number density of the dark matter particle divided by the critical density of the Universe. The present number density $n_{\chi}$ of the relic particle is obtained by numerically solving the Boltzmann equation
\begin{equation}
	\frac{{\rm d}n_{\chi}}{{\rm d}t} ~=~ -3 H n_{\chi} - \langle \sigma_{\rm ann}v \rangle \big( n_{\chi}^2 - n_{\rm eq}^2 \big) \,.
\end{equation}
Here, the Hubble parameter $H$ translates the expansion of the Universe, while particle physics enters through the thermally averaged annihilation cross section $\sigma_{\rm ann}$. A comparison of the predicted relic abundance to the experimentally value obtained by Planck \cite{Planck2015},
\begin{equation}
	\Omega_{\rm CDM}h^2 ~=~ 0.1199 \pm 0.0022 \,,
	\label{Eq:OmegaCDM}
\end{equation}
allows to identify (dis)favoured regions of the new physics parameter space. In recent years, this measurement has become more and more precise, the current precision reaching about 2\%. It is therefore necessary that the theoretical prediction meets this precision. The actual calculation of $\Omega_{\chi}h^2$ is typically carried out by public computer codes such as \texttt{micrOMEGAs} \cite{micrOMEGAs} and \texttt{DarkSUSY} \cite{DarkSUSY}. All relevant (co)annihilation processes are implemented in these codes, but only at tree-level. However, the need of an increase in precision clearly calls for higher-order corrections to $\sigma_{\rm ann}$. 

The project \texttt{DM@NLO} (``Dark Matter at Next-to-Leading-Order'') aims at providing a calculation of $\sigma_{\rm ann}$ including radiative corrections of order $\alpha_s$ in QCD. In the following, we focus on the Minimal Supersymmetric Standard Model (MSSM) and assume the lightest neutralino $\tilde{\chi}^0_1$ to be the dark matter candidate. The annihilation cross-section then comprises neutralino pair annihilation, its coannihilation with a sfermion or another gaugino, as well as pair annihilations of sfermions and other gauginos. The relative weights of these processes depend on the parameter set under investigation. 

In this contribution, we briefly review results for selected (co)annihilation processes treated within \texttt{DM@NLO}. In the second part, we focus on the evaluation of the theory uncertainty, which has become possible for the first time thanks to the inclusion of radiative corrections in the calculation of the annihilation cross-section and the relic density. Finally, perspectives are discussed.

\section{Review of selected results}

In the MSSM, most of the processes given above receive radiative corrections in QCD since they involve coloured particles in either the initial or the initial state. The relevant one-loop contributions are self-energy, vertex, and box diagrams. These have to be combined with the real emission of a gluon, which is a correction of the same order and ensures infrared finiteness. At present, the processes covered by the \texttt{DM@NLO} project are \cite{DMNLOChiChi, DMNLOCoAnn, DMNLO2014Stop, DMNLO2016Scale}
\begin{eqnarray}
	\tilde{\chi} \tilde{\chi} ~\to~ q \bar{q}' \,, \quad
	\tilde{\chi} \tilde{q} ~\to~ q H / q V / q g \,, \quad
	\tilde{q} \tilde{q}^* ~\to~ HH / VV / HV \,,
	\label{Eq:Processes1}
\end{eqnarray}
including gauginos $\tilde{\chi} = \{ \tilde{\chi}^0_{1,2,3,4}, ~\tilde{\chi}^{\pm}_{1,2} \}$, (s)quarks with $q=\{u,d,c,s,t,b \}$, Higgs bosons $H=\{h^0, H^0, A^0, H^{\pm}\}$, electroweak vector bosons $V=\{\gamma, Z^0, W^{\pm} \}$, and gluons $g$. These processes have been calculated analytically and implemented in a numerical code, which provides the calculation of the annihilation cross-section $\sigma_{\rm ann}$ including the QCD corrections. On the technical side, we have defined a dedicated renormalization scheme, which is applicable to all the above classes of processes. Moroever, in order to obtain infrared finite results, we apply either the phase space slicing or the dipole subtraction method. In the case of stop pair annihilation, our calculation also includes the resummation of Coulomb corrections. Currently, \texttt{DM@NLO} is used with \texttt{micrOMEGAs}, while an interface to \texttt{DarkSUSY} is in development.

\begin{figure}[t]
	\begin{center}
		\includegraphics[scale=0.45]{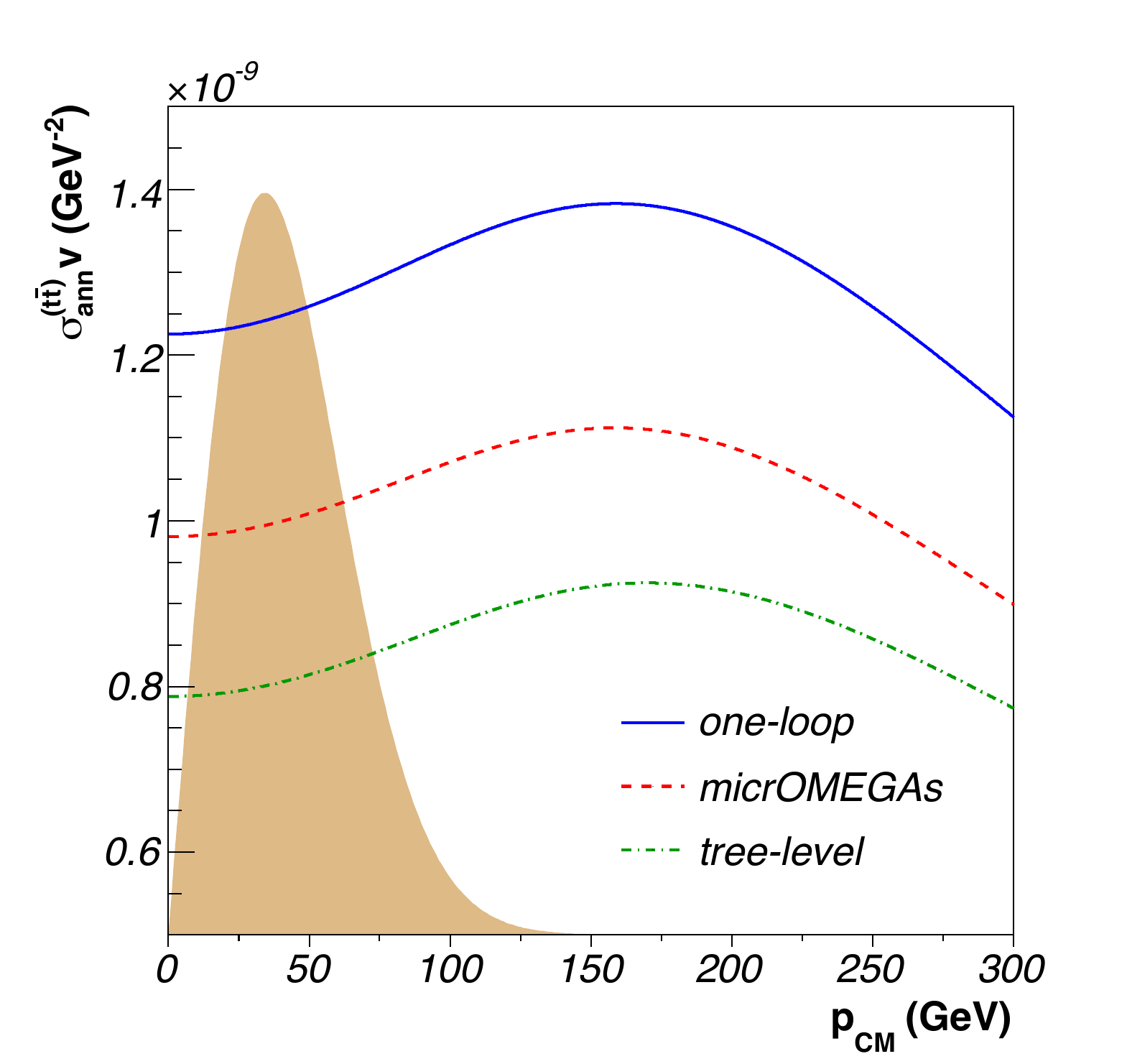}
		\includegraphics[scale=0.45]{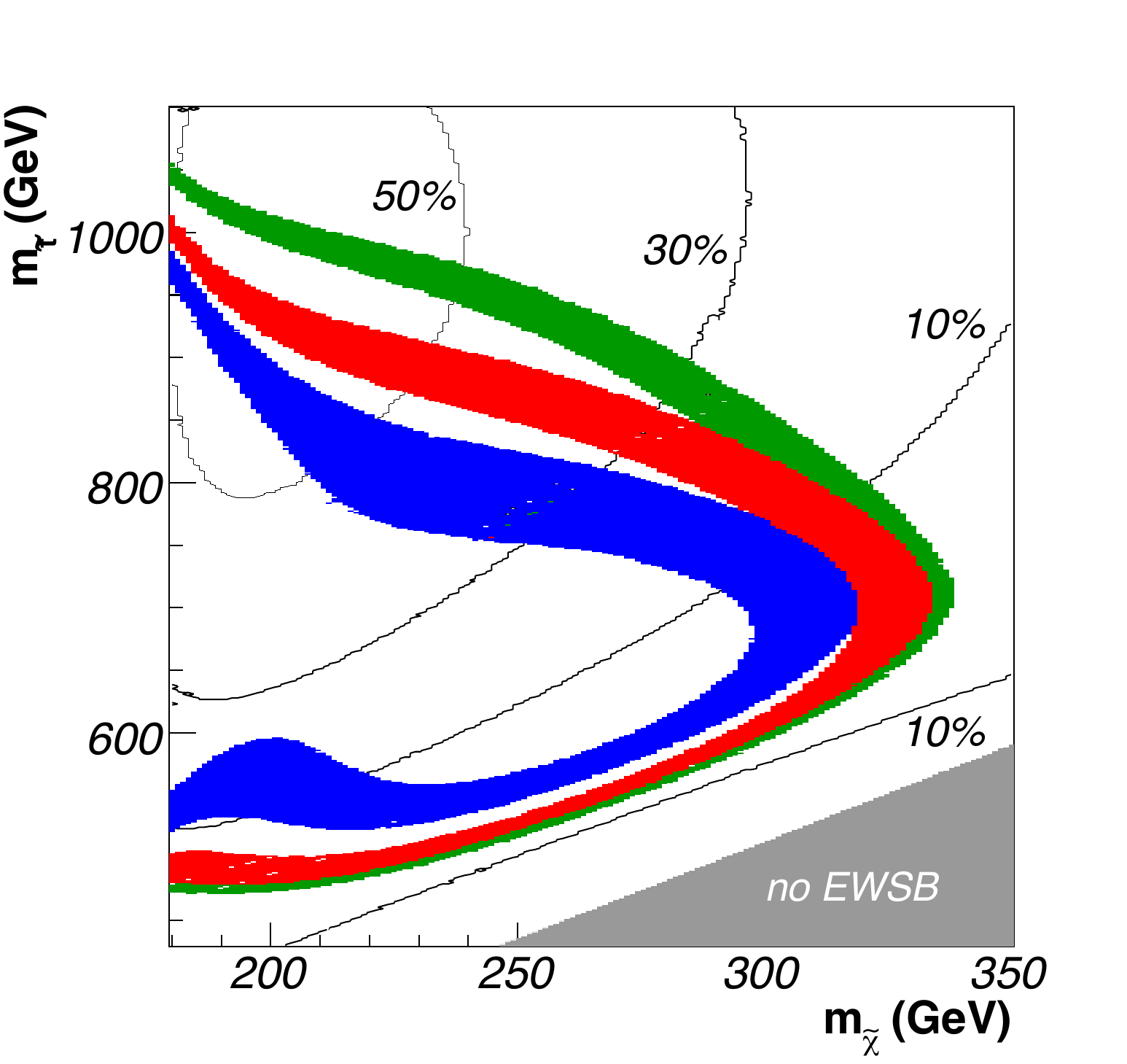}
	\end{center}
	\vspace*{-7mm}
	\caption{Left: Annihilation cross-section of the process $\tilde{\chi}^0_1 \tilde{\chi}^0_1 \to t\bar{t}$ as a function of the centre-of-momentum energy obtained by \texttt{DM@NLO} at tree-level and one-loop level, together with the \texttt{micrOMEGAs} result. The shaded area indicates the velocity distribution of the neutralino at the time of freeze-out (in arbitrary units). Right: Resulting favoured regions with respect to the neutralino relic density in the $m_{\tilde{\chi}^0_1}$-$m_{\tilde{t}_1}$ plane, obtained using \texttt{DM@NLO} (tree-level and one-loop level) and \texttt{micrOMEGAs}.}
	\label{Fig:Results1}
\end{figure}

\begin{figure}[t]
	\begin{center}
		\includegraphics[scale=0.36]{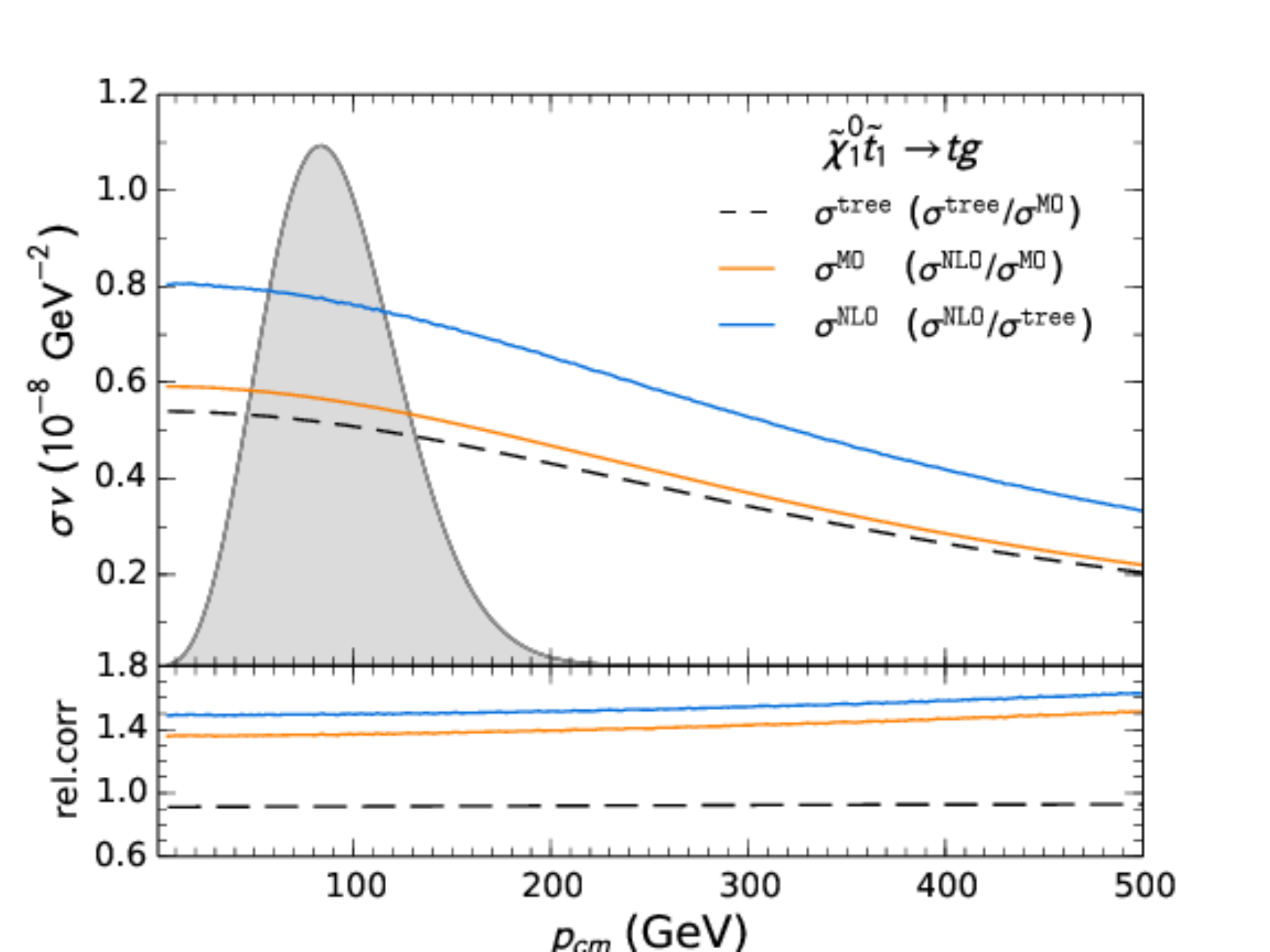}
		\includegraphics[scale=0.36]{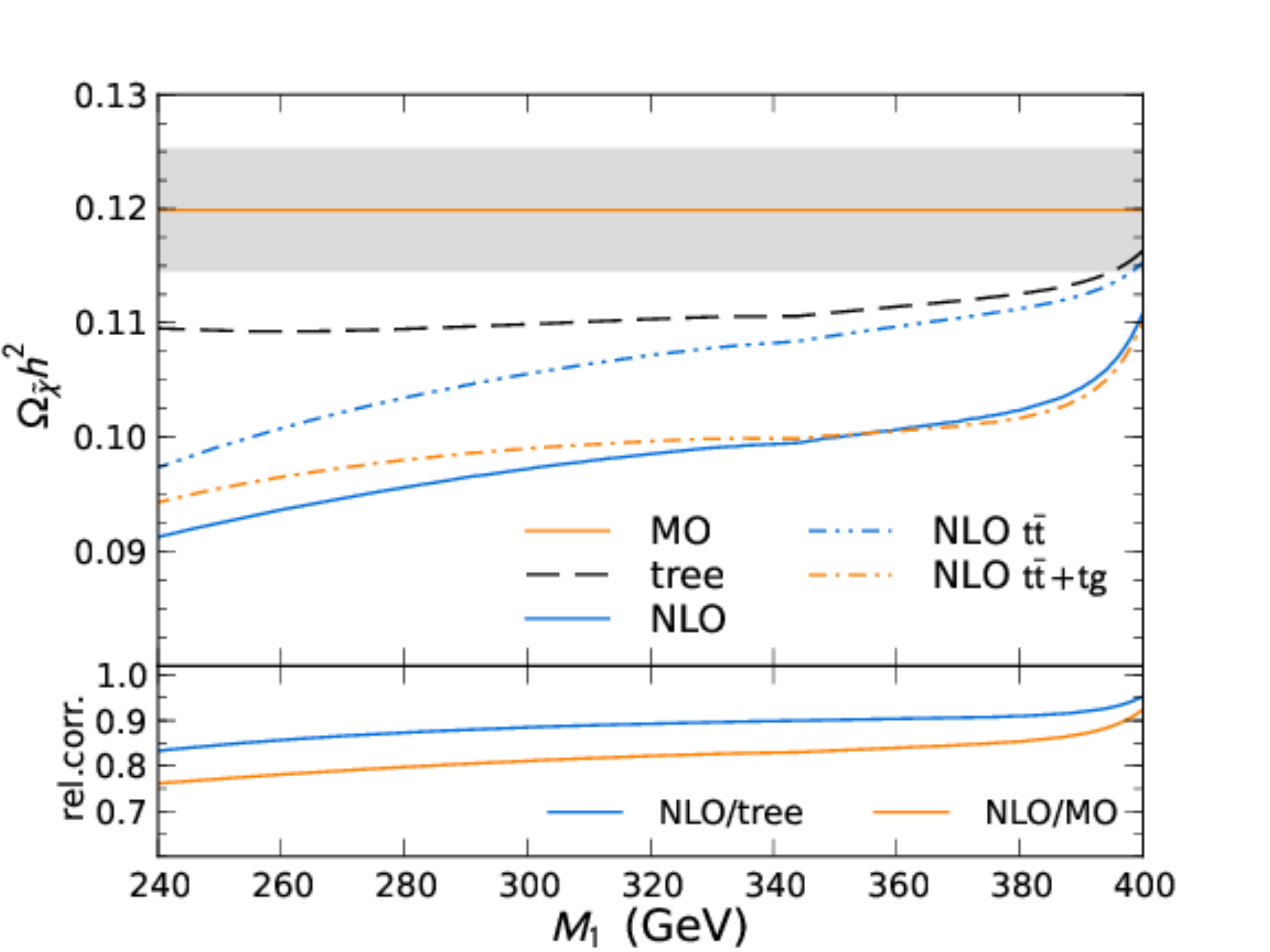}
	\end{center}
	\vspace*{-5mm}
	\caption{Left: Same as Fig.\ 1 (left) for the coannihilation process $\tilde{\chi}^0_1 \tilde{t}_1 \to tg$. The lower part shows in addition the ratios of the different values. Right: Neutralino relic density as a function of the bino mass parameter $M_1$. We show the values obtained using the \texttt{DM@NLO} calculations as well as the \texttt{micrOMEGAs} result.}
	\label{Fig:Results2}
\end{figure}

\begin{figure}[t]
	\begin{center}
		\includegraphics[scale=0.36]{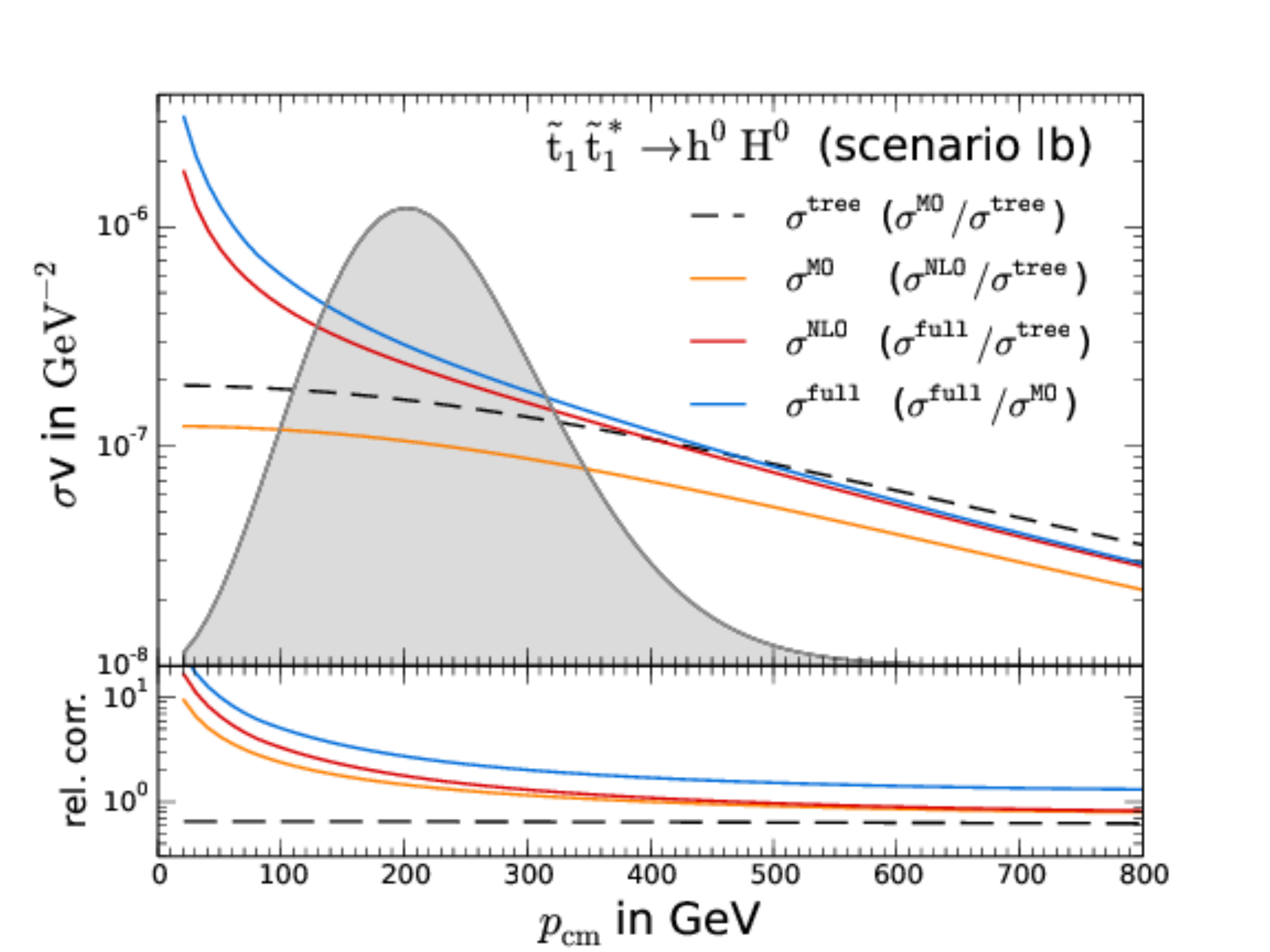}
		\includegraphics[scale=0.36]{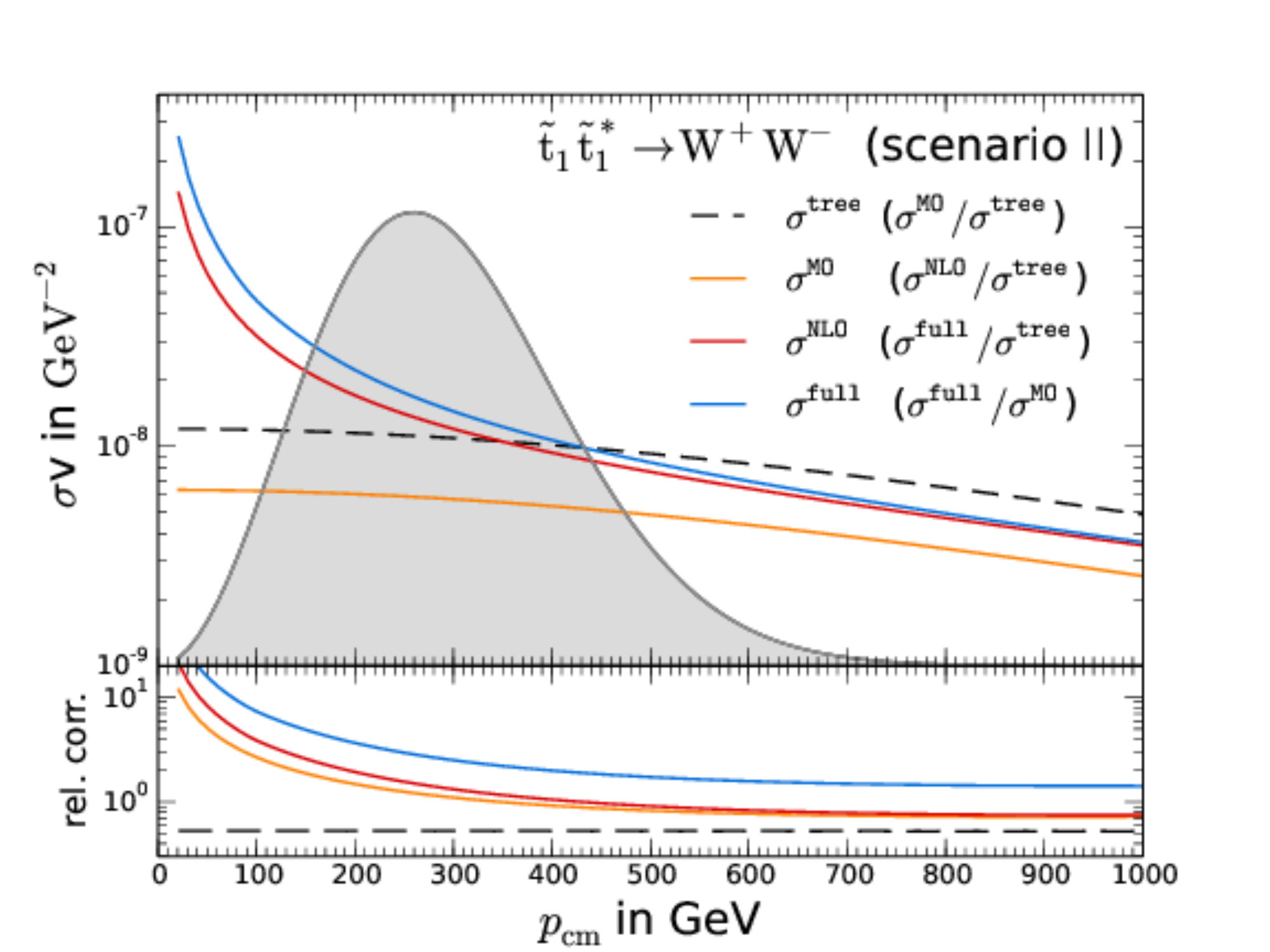}
	\end{center}
	\vspace*{-5mm}
	\caption{Same as Fig.\ 2 (left) for two examples of stop-antistop annihilation. We show in addition the result obtained when taking into account the Coulomb resummation ($\sigma^{\rm full}$).}
	\label{Fig:Results3}
\end{figure}

In Figs.\ \ref{Fig:Results1}, \ref{Fig:Results2}, and \ref{Fig:Results3} we show the results for three example processes, evaluated for parameter sets which lead to a dominant contribution of the respective process. As can be seen in all cases, the radiative corrections can increase the annihilation cross-section by up to about 50\%. This reflects in a correction to the relic density, which can be more important than the uncertainty given in Eq.\ (\ref{Eq:OmegaCDM}). Consequently, these corrections should be taken into account when analysing the MSSM parameter space or when extracting supersymmetric parameters from cosmological measurements. 

Concerning the annihilation of a neutralino pair into quarks \cite{DMNLOChiChi}, shown in Fig.\ \ref{Fig:Results1}, let us note that the effective Yukawa couplings included in \texttt{micrOMEGAs} are a very good approximation in the direct vicinity of a Higgs resonance, in particular for the process $\tilde{\chi}^0_1 \tilde{\chi}^0_1 \to b\bar{b}$. However, other subchannels can be dominant, e.g.\ the exchange of a $Z^0$ or a squark, and the corresponding radiative corrections cannot be covered by effective tree-level couplings. Moreover, the Higgs resonance may not coincide with the peak of the thermal velocity distribution of the dark matter particle. 

Coming to the coannihilation of a neutralino with a stop \cite{DMNLOCoAnn}, see Fig.\ \ref{Fig:Results2}, note that the dominant final states are a top with a gluon or a Standard-Model Higgs boson. While the former is important because of phase-space reasons and the strong coupling involved, the latter is favoured by light stops and sizeable trilinear couplings, which are needed to satisfy the measured Higgs-boson mass of about 125 GeV, see Refs.\ \cite{DMNLOCoAnn} for details on the different subchannels.

Finally, in case of the stop-antistop annihilation \cite{DMNLO2014Stop}, shown in Fig.\ \ref{Fig:Results3}, Coulomb corrections are relevant in the low-velocity region, while in the high-velocity region the fixed-order corrections are dominant.

All examples show that the impact of the NLO corrections can be numerically more important than the current uncertainty of the Planck results given in Eq.\ (\ref{Eq:OmegaCDM}). 

\section{Evaluation of the theory error}

The computation of one-loop corrections not only allows to obtain more precise predictions for the (co)annihilation cross-section and the relic density of the neutralino. It also allows to estimate the error on the theoretical predictions. More precisely, a fixed-order calculation is affected by the presence of an unphysical parameter, the renormalization scale $\mu_R$, which allows for some freedom in the choice of its value. Generally speaking, the error introduced in an NLO calculation is of NNLO order, in the same way as also the uncertainty related to the choice of the renormalization scheme. In the calculation leading to the results presented above, the renormalization scale has been identified with the scale at which the supersymmetric mass spectrum and associated parameters are input to \texttt{micrOMEGAs} and our \texttt{DM@NLO} code. By default, this is done at $\mu_R = 1$ TeV according to the SPA convention \cite{SPA}. The variation of $\mu_R$ by a factor of two around this central value gives an estimation of the theoretical uncertainty which affects the calculation. 

\begin{figure}[t]
	\begin{center}
		\includegraphics[scale=0.38]{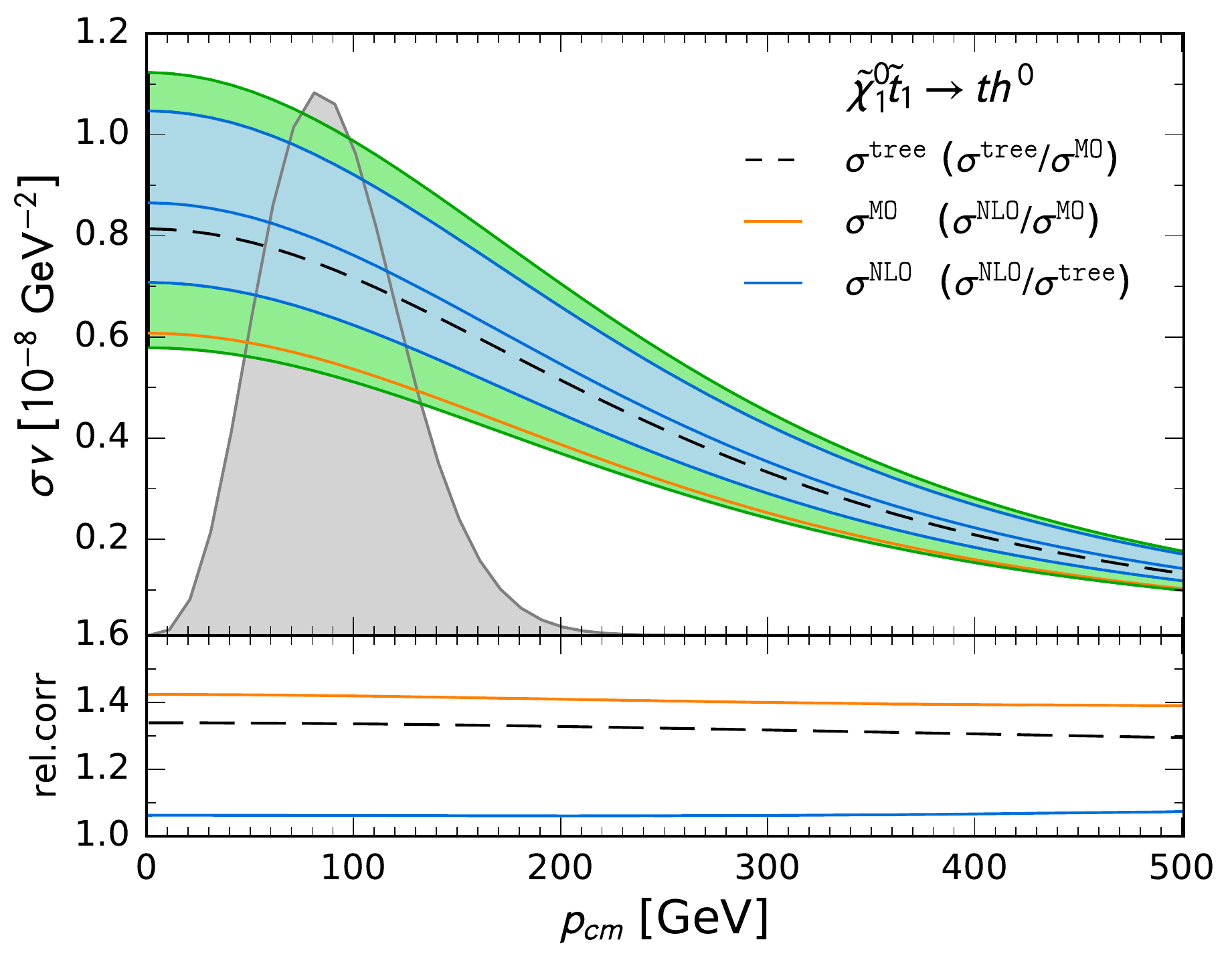}
		\quad
		\includegraphics[scale=0.38]{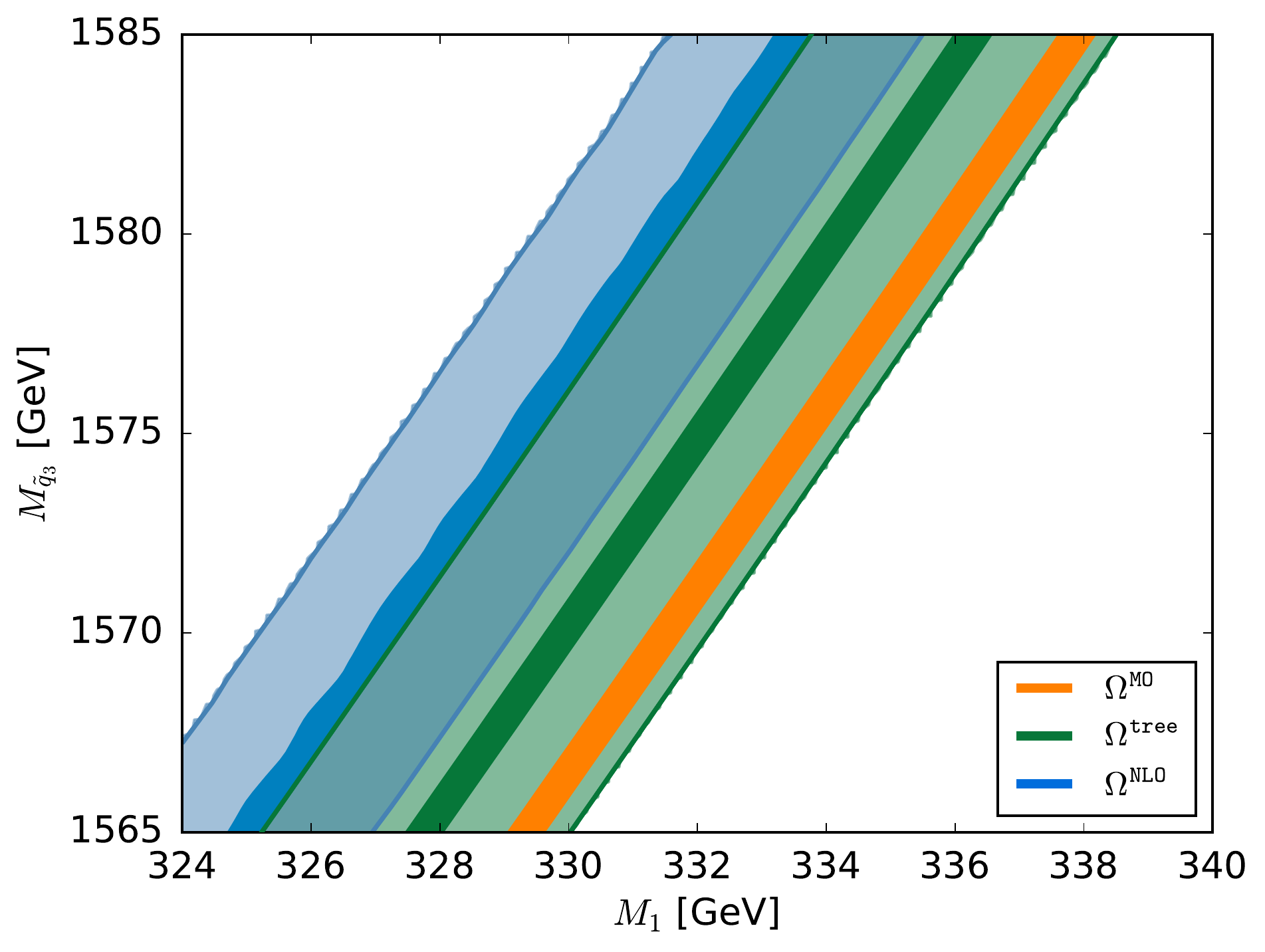}
	\end{center}
	\vspace*{-5mm}
	\caption{Left: Same as Fig.\ 2 (left) for the process $\tilde{\chi}^0_1 \tilde{t}_1 \to t h^0$. Right: Same as Fig.\ 1 (right) in the $M_1$-$M_{\tilde{q}_3}$ plane. In both graphs, we indicate the error bands corresponding to the variation of the renormalization scale for the \texttt{DM@NLO} tree-level and NLO results.}
	\label{Fig:ScaleVar}
\end{figure}

In the first panel of Fig.\ \ref{Fig:ScaleVar} we show as an example the results for the coannihilation cross-section of the process $\tilde{\chi}^0_1 \tilde{t}_1 \to t h^0$. The error corresponding to a variation of the scale between $\mu_R=0.5$ and $\mu_R=2$ TeV is indicated by the bands around our tree-level and NLO results. 

The first observation to be made is that the \texttt{micrOMEGAs} result lies within the error band associated to our tree-level calculation. This is expected since the difference originates from the use of different schemes, see Refs.\ \cite{DMNLOCoAnn, DMNLOChiChi} for details. Second, we observe that the uncertainty is reduced when going from the tree-level to the NLO calculation. Again, this is expected since the difference between two schemes is expected to be of higher order. Finally, we see that the \texttt{micrOMEGAs} result lies outside the band associated to the NLO calculation, underlining the importance of taking into account the radiative corrections.

The same analysis on the level of the neutralino relic density, shown in the right panel of Fig.\ \ref{Fig:ScaleVar}, leads to the same conclusions.

\section{Conclusion and perspectives}

As discussed above, radiative corrections to dark matter annihilation are important since they lead to more precise predictions of the relic density, which is often invoked in the study of new physics models. We have shown that in case of neutralino dark matter, the impact of corrections of order $\alpha_s$ in QCD can be more important than the current experimental uncertainty in wide regions of the MSSM parameter space. These corrections should therefore be taken into account in global studies of the model or when extracting mass parameters from cosmological measurements. 

While the processes given in (\ref{Eq:Processes1}) are completely implemented, additional classes of processes receive QCD corrections and are yet to be implemented. Using the same generic notations as above, these processes are
\begin{eqnarray}
	\tilde{\ell} \tilde{\ell}^* ~\to~ q \bar{q}' \,, \quad 
	\tilde{\chi} \tilde{\chi} ~\to~ \gamma\gamma / gg \,, \quad 
	\tilde{q} \tilde{q}^* ~\to~ q \bar{q} / gg \,, \quad
	\tilde{q} \tilde{q} ~\to~ qq \,,
\end{eqnarray}
and are subject to current work. 

For more detailed discussions, presentations and discussions of the obtained results, the interested reader is referred to Refs.\ \cite{DMNLOChiChi, DMNLOCoAnn, DMNLO2014Stop, DMNLO2016Scale}. Recently, radiative corrections to the neutralino-nucleon cross-section, relevant for direct detection of dark matter, have been presented \cite{DMNLO2016DD}. 


\end{document}